\documentclass[acmsmall]{acmart}

\usepackage{balance}  
\usepackage{graphicx}
\usepackage{amsfonts}
\usepackage{url}
\usepackage{algorithm}
\usepackage{algorithmic}
\usepackage{float}
\usepackage{amssymb,amsmath,epsfig,graphics,enumerate,float,subfigure}
\usepackage{graphicx}
\usepackage{color}
\usepackage{xspace}
\usepackage{multirow}

\newcommand{\kw}[1]{{\ensuremath {\mathsf{#1}}}\xspace}
\newcommand{\kwnospace}[1]{{\ensuremath {\mathsf{#1}}}}

\newcommand{\stitle}[1]{\vspace{1ex} \noindent{\bf #1}}

\newcommand{\NPhard}{NP-hard\xspace}

\newcommand{\icgs}{\kw{ICG}-\kw{S}}
\newcommand{\icgm}{\kw{ICG}-\kw{M}}
\newcommand{\leks}{\kw{LEKS}}

\newcommand{\MST}{\kw{MST}}
\newcommand{\MWP}{\kw{MWP}}
\newcommand{\dist}{\kw{dist}}
\newcommand{\spath}{\kwnospace{spath}}
\newcommand{\ltree}{\kw{LEKS}-\kw{tree}}
\newcommand{\lpath}{\kw{LEKS}-\kw{path}}

\newcommand{\squishlisttight}{
 \begin{list}{$\bullet$}
  { \setlength{\itemsep}{0pt}
    \setlength{\parsep}{0pt}
    \setlength{\topsep}{0pt}
    \setlength{\partopsep}{0pt}
    \setlength{\leftmargin}{2em}
    \setlength{\labelwidth}{1.5em}
    \setlength{\labelsep}{0.5em}
} }

\newcommand{\squishnumlist} {
\newcounter{qcounter}
\begin{list}{\arabic{qcounter}.~}{\usecounter{qcounter}} 
{  \setlength{\itemsep}{0pt}
    \setlength{\parsep}{0pt}
    \setlength{\topsep}{0pt}
    \setlength{\partopsep}{0pt}
    \setlength{\leftmargin}{2em}
    \setlength{\labelwidth}{1.5em}
    \setlength{\labelsep}{0.5em}
}}

\newcommand{\squishend}{
  \end{list}
}

\begin{document}

\title{Fast Algorithms for Intimate-Core Group Search in Weighted Graphs}

\author{Longxu Sun}

\affiliation{%
  \institution{Hong Kong Baptist University}
  \city{Hong Kong}
  \country{China}}
\email{sunlongxu@life.hkbu.edu.hk}

\author{Xin Huang}
\affiliation{%
  \institution{Hong Kong Baptist University}
  \city{Hong Kong}
  \country{China}}
\email{xinhuang@comp.hkbu.edu.hk}

\author{Rong-Hua Li}

\affiliation{%
  \institution{Beijing Institute of Technology}
  \city{Beijing}
  \country{China}}
\email{lironghuabit@126.com}

\author{Jianliang Xu}

\affiliation{%
  \institution{Hong Kong Baptist University}
  \city{Hong Kong}
  \country{China}}
\email{xujl@comp.hkbu.edu.hk}

\renewcommand{\shortauthors}{Sun et al.}

\begin{abstract}

Community search that finds query-dependent communities has been studied on various kinds of graphs. As one instance of community search, intimate-core group search over a weighted graph is to find a connected $k$-core containing all query nodes with the smallest group weight. However, existing state-of-the-art methods start from the maximal $k$-core to refine an answer, which is practically inefficient for large networks. In this paper, we develop an efficient framework, called \underline{l}ocal \underline{e}xploration \underline{k}-core \underline{s}earch (LEKS), to find intimate-core groups in graphs. We propose a small-weighted spanning tree to connect query nodes, and then expand the tree level by level to a connected $k$-core, which is finally refined as an intimate-core group. We also design a protection mechanism for critical nodes to avoid the collapsed $k$-core. Extensive experiments on real-life networks validate the effectiveness and efficiency of our methods.


\end{abstract}

\maketitle

\section{Introduction}

Graphs widely exist in social networks, biomolecular structures, traffic networks, world wide web, and so on. Weighted graphs have not only the simple topological structure but also edge weights. The edge weight is often used to indicate the strength of the relationship, such as interval in social communications, traffic flow in the transportation network, carbon flow in the food chain, and so on~\cite{newman2001scientific,opsahl2010node,newman2004analysis}. Weighted graphs provide information that better describes the organization and hierarchy of the network, which is helpful for community detection~\cite{newman2004analysis} and community search~\cite{hlx2019community,yuan2017index,fang2019survey,huang2014querying}. Community detection aims at finding all communities on the entire network, which has been studied a lot in the literature. Different from community detection, the task of community search finds only query-dependent communities, which has a wide application of disease infection control, tag recommendation, and social event organization~\cite{sozio2010community,zheng2017querying}. Recently, several community search models have been proposed in different dense subgraphs of $k$-core~\cite{batagelj2003m,sariyuce2013streaming} and $k$-truss~\cite{wang2012truss,huang2014querying}.  

\begin{figure}[ht]
     \begin{tabular} {@{\hspace{0.1\textwidth}}c @{\hspace{0.06\textwidth}} c}
     	\includegraphics[width=0.50\textwidth]{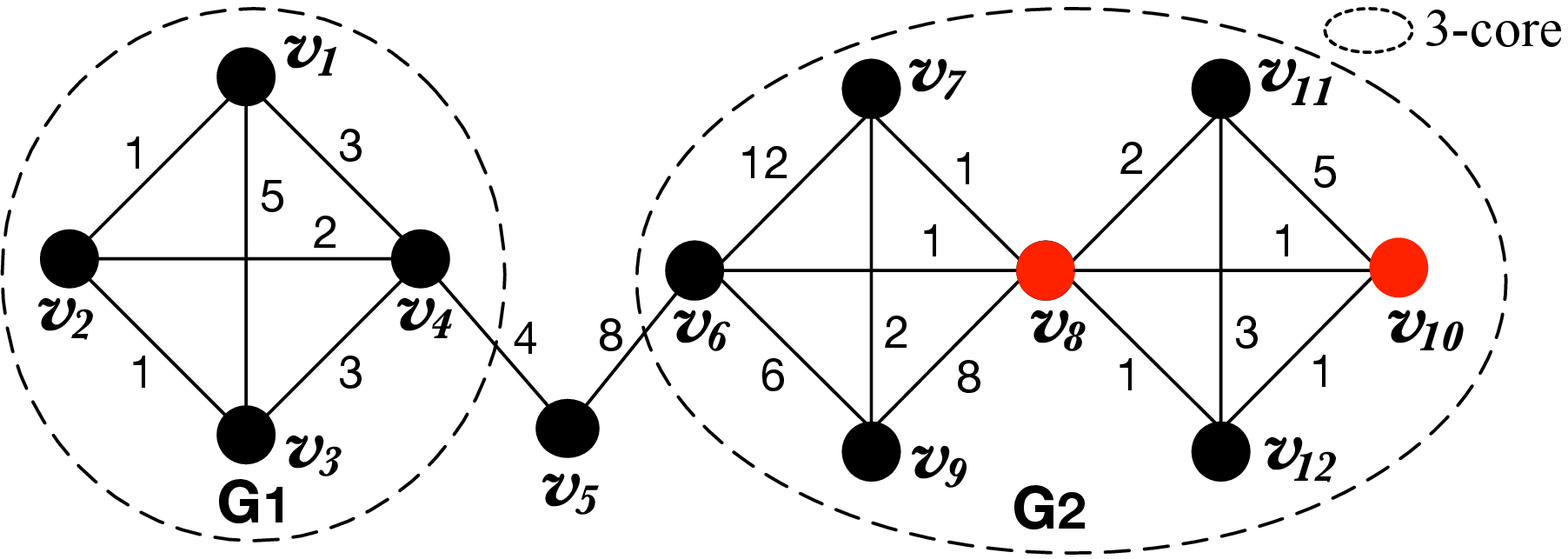} &
     	\includegraphics[width=0.15\textwidth]{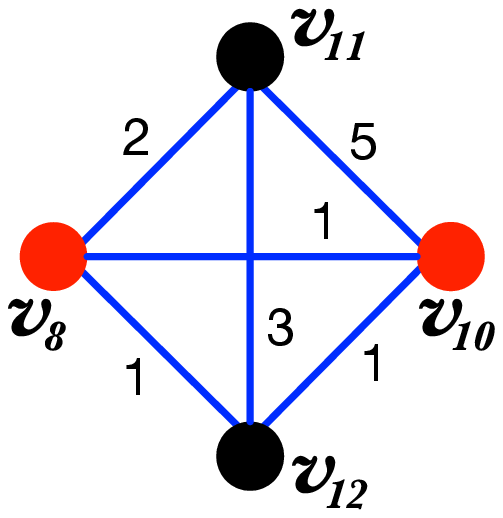}\\
     	\newline
     	(a) Graph $G$ & (b) Intimate-core group\\
     \end{tabular}
     \caption{An example of intimate-core group search in graph $G$ for $Q=\{v_8, v_{10}\}$ and $k=3$.}
     \label{fig.motivate}
\end{figure}

As a notation of dense subgraph, $k$-core requires that every vertex has $k$ neighbors in the $k$-core. For example, Fig.~\ref{fig.motivate}(a) shows a graph $G$. Subgraphs $G_1$ and $G_2$ are both connected 3-cores, in which each vertex has at least three neighbors. $K$-core has been popularly used in many community search models~\cite{zhu2018k,fang2016effective,li2015influential,sozio2010community,barbieri2015efficient,zhu2018k,medya2019k}. Recently, Zheng et al.~\cite{zheng2017querying} proposed one problem of intimate-core group search in weighted graphs as follows. 

\stitle{Motivating example.} Consider a social network $G$ in Fig.~\ref{fig.motivate}(a). Two individuals have a closer friendship if they have a shorter interval for communication, indicating a smaller weight of the relationship edge. The problem of intimate-core group search aims at finding a densely-connected $k$-core containing query nodes $Q$ with the smallest group weight as an answer. For $Q=\{v_{8}, v_{10}\}$ and $k=3$, the intimate-core group is shown in Fig.~\ref{fig.motivate}(b) with a minimum group weight of 13.

This paper studies the problem of intimate-core group search in weighted graphs. Given an input of query nodes in a graph and a number $k$, the problem is to find a connected $k$-core containing query nodes with the smallest weight. In the literature, existing solutions proposed in~\cite{zheng2017querying} find the maximal connected $k$-core and iteratively remove a node from this subgraph for intimate-core group refinement. However, this approach may take a large number of iterations, which is inefficient for big graphs with a large component of $k$-core. Therefore, we propose a solution of local exploration to find a small candidate $k$-core, which takes a few iterations to find answers. To further speed up the efficiency, we build a $k$-core index, which keeps the structural information of $k$-core for fast identification. Based on the $k$-core index, we develop a local exploration algorithm \leks for intimate-core group search.  Our algorithm \leks first generates a tree to connect all query nodes, and then expands it to a connected subgraph of $k$-core. Finally, \leks keeps refining candidate graphs into an intimate-core group with small weights. We propose several well-designed strategies for \leks to ensure the fast-efficiency and high-quality of answer generations.  

\stitle{Contributions.} Our main contributions of this paper are summarized as follows. 

\begin{itemize}
\item We investigate and tackle the problem of intimate-core group search in weighted graphs, which has wide applications on real-world networks.
The problem is \NPhard, which bring challenges to develop efficient algorithms. 

\item We develop an efficient local exploration framework of \leks based on the $k$-core index for intimate-core group search. \leks consists of three phases: tree generation, tree-to-graph expansion, and intimate-core refinement.

\item  In the phase of tree generation, we propose to find a seed tree to connect all query nodes, based on two generated strategies of \emph{spanning tree} and \emph{weighted path} respectively. Next, we develop the tree-to-graph expansion, which constructs a hierarchical structure by expanding a tree to a connected $k$-core subgraph level by level. Finally, we refine a candidate $k$-core to an intimate-core group with a small weight. During the phases of expansion and refinement, we design a protection mechanism for query nodes, which protects critical nodes to collapse the $k$-core. 


\item Our experimental evaluation demonstrates the effectiveness and efficiency of our \leks algorithm on real-world weighted graphs. We show the superiority of our methods in finding intimate groups with smaller weights, against the state-of-the-art \icgm method~\cite{zheng2017querying}.
\end{itemize}


\stitle{Roadmap.} The rest of the paper is organized as follows. Section~\ref{sec.relate} reviews the previous work related to ours. Section~\ref{sec.problem} presents the basic concepts and formally defines our problem. Section~\ref{sec.algo} introduces our index-based local exploration approach \leks.  Section~\ref{sec.exp} presents the experimental evaluation. Finally, Section~\ref{sec.con} concludes the paper. 

\section{Related Work} \label{sec.relate}

In the literature, numerous studies have been investigated community search based on various kinds of dense subgraphs, such as  $k$-core~\cite{batagelj2003m,sariyuce2013streaming}, $k$-truss~\cite{wang2012truss,huang2014querying} and clique~\cite{yuan2017index,yuan2016diversified}. Community search has been also studied on many labeled graphs, including weighted graphs~\cite{duan2009community,zheng2017finding,zheng2017querying}, influential graphs~\cite{li2015influential,bi2018optimal}, and keyword-based graphs~\cite{fang2017effective,fang2016effective,huang2017attribute}. 
Table~\ref{tab.related} compares different characteristics of existing community search studies and ours.

\vspace{5 pt}

\begin{table}[ht] 
\centering
\small
\caption{A comparison of existing community search studies and ours}\label{tab.related}
\begin{tabular}{|c|c|c|c|c|c|c|c|}
\hline
\multirow{2}*{Method} & Dense Subgraph & Node & Edge & Local & \multirow{2}*{Index-based} & Multiple & \multirow{2}*{\NPhard}\\
	~& Model & Type & Type & Search &~	& Query Nodes &~\\
\hline
\cite{yuan2016diversified} & clique & $\times$ & $\times$ & \checkmark & \checkmark & $\times$ & \checkmark\\
\cite{yuan2017index} & clique & $\times$ & $\times$ & $\times$ & \checkmark & \checkmark & \checkmark\\
\cite{huang2015approximate} & $k$-truss & $\times$ & $\times$ & \checkmark & \checkmark & \checkmark & \checkmark\\
\cite{zhu2018k,medya2019k} & $k$-core & $\times$ & $\times$ & $\times$ & $\times$ & $\times$ & \checkmark\\
\cite{cui2014local} & $k$-core & $\times$ & $\times$ & \checkmark & $\times$ & $\times$ & \checkmark\\
\cite{sozio2010community} & $k$-core & $\times$ & $\times$ & $\times$ & $\times$ & \checkmark & \checkmark\\
\cite{barbieri2015efficient} & $k$-core & $\times$ & $\times$ & \checkmark & \checkmark & \checkmark & \checkmark\\
\cite{huang2017attribute} & $k$-truss & keyword & $\times$ & \checkmark & \checkmark & \checkmark & \checkmark\\
\cite{fang2016effective} & $k$-core & keyword & $\times$ & \checkmark & \checkmark & $\times$  & \checkmark\\
\cite{li2015influential} & $k$-core & influential & $\times$ & $\times$ & \checkmark & $\times$  & $\times$\\
\cite{bi2018optimal} & $k$-core & influential & $\times$ & \checkmark & $\times$ & $\times$  & $\times$\\
\cite{zheng2017finding} & $k$-truss & $\times$ & weighted & \checkmark & \checkmark & $\times$ & $\times$\\
 \cite{zheng2017querying} & $k$-core & $\times$ & weighted & $\times$ & $\times$ & \checkmark & \checkmark\\
Ours & $k$-core & $\times$ & weighted & \checkmark & \checkmark & \checkmark & \checkmark\\
\hline
\end{tabular}
\end{table}

The problem of $k$-core minimization~\cite{barbieri2015efficient,zhu2018k,medya2019k,cui2014local} aims to find a minimal connected $k$-core subgraph containing query nodes. The minimum wiener connector problem is finding a small connected subgraph to minimize the sum of all pairwise shortest-path distances between the discovered vertices~\cite{ruchansky2015minimum}. Different from all the above studies, our work aims at finding an intimate-core group containing multiple query nodes in weighted graphs. We propose fast algorithms for intimate-core group search, which outperform the state-of-the-art method~\cite{zheng2017querying} in terms of quality and efficiency. 


%
%
\section{Preliminaries} \label{sec.problem}

In this section, we formally define the problem of intimate-core group search and revisit the existing intimate-core group search approaches.


\subsection{Problem Definition}
Let $G(V, E, w)$ be a weighted and undirected graph where $V$ is the set of nodes, $E$ is the set of edge, and $w$ is an edge weight function. Let $w(e)$ to indicate the weight of an edge $e\in E$. 
The number of nodes in $G$ is defined as $n = |V|$.  The number of edges in $G$ is defined as $m =|E|$. We denote the set of neighbors of a node $v$ by $N_{G}(v) = \{u\in V: (u,v)\in E\}$, and the degree of $v$ by $deg_{G}(v)=|N_{G}(v)|$. For example, Fig.~\ref{fig.motivate}(a) shows a weighted graph $G$. Node $v_{5}$ has two neighbors as $N_{G}(v_{5}) = \{v_{4}, v_{6}\}$, thus the degree of $v_{5}$ is $deg_{G}(v_{5})=2$ in graph $G$. Edge $(v_2, v_3)$ has a weight of $w(v_2, v_3) =1$. Based on the definition of degree, we can define the $k$-core as follows. 

\begin{definition}
[K-Core~\cite{batagelj2003m}] Given a graph $G$, the $k$-core is the largest subgraph $H$ of $G$ such that every node $v$ has degree at least $k$ in $H$, i.e., $deg_{H}(v)\geq k$.
\end{definition}

For a given integer $k$, the $k$-core of graph $G$ is denoted by $C_{k}(G)$, which is determinative and unique by the definition of largest subgraph constraint. For example, the 3-core of $G$ in Fig.~\ref{fig.motivate}(a) has two components $G_1$ and $G_2$. Every node has at least 3 neighbors in $G_1$ and $G_2$ respectively. However, the nodes are disconnected between $G_1$ and $G_2$ in the 3-core  $C_3(G)$. To incorporate connectivity into $k$-core, we define a connected $k$-core.    

\begin{definition}
[Connected K-Core] Given graph $G$ and number $k$, a connected $k$-core $H$ is a connected component of $G$ such that every node $v$ has degree at least k in $H$, i.e., $deg_H(v)\geq k$.
\end{definition}

Intuitively, all nodes are reachable in a connected $k$-core, i.e., there exist paths between any pair of nodes. $G_1$ and $G_2$ are two connected 3-cores  in Fig.~\ref{fig.motivate}(a). 




\begin{definition}
	[Group Weight] Given a subgraph $H \subseteq G$, the group weight of $H$, denoted by $w(H)$, is defined as the sum of all edge weights in $H$, i.e., $w(H) = \sum_{e\in E(H)} w(e)$.
\end{definition}

\begin{example}
	For the subgraph $G_1 \subseteq G$ in Fig.~\ref{fig.motivate}(a), the group weight of $G_1$ is $w(G_1)= \sum_{e\in E(G_1)} w(e)= 1+3+5+2+1+3 = 15$.
\end{example}

On the basis of the definitions of connected $k$-core and group weight, we define the \emph{intimate-core group} in a graph $G$ as follows. 

\begin{definition}
	[Intimate-Core Group~\cite{zheng2017querying}] Given a weighted graph $G = (V, E, w)$, a set of query nodes $Q$ and a number $k$, the intimate-core group is a subgraph $H$ of $G$ if $H$ satisfies following conditions:
\begin{itemize}
	\item \textbf{Participation.} $H$ contains all the query nodes $Q$, i.e., $Q \subseteq V_{H}$;
	\item \textbf{Connected K-Core.} $H$ is a connected $k$-core with $deg_{H}(v)\geq k$;
	\item \textbf{Smallest Group Weight.} The group weight $w(H)$ is the smallest, that is, there exists no $H' \subseteq G$ achieving a group weight of $w(H^{'}) < w(H)$ such that $H^{'}$ also satisfies the above two conditions.
\end{itemize}
\end{definition}

Condition (1) of participation makes sure that the intimate-core group contains all query nodes. Moreover, Condition (2) of connected $k$-core requires that all group members are densely connected with at least $k$ intimate neighbors. In addition, Condition (3) of minimized group weight ensures that the group has the smallest group weight, indicating the most intimate in any kinds of edge semantics. A small edge weight means a high intimacy among the group. Overall, intimate core groups have several significant advantages of small-sized group, offering personalized search for different queries, and close relationships with strong connections.

The problem of \emph{intimate-core group search} studies in this paper is formulated in the following.

\stitle{Problem formulation: } Given an undirected weighted graph $G(V, E, w)$, a number $k$, and a set of query nodes $Q$, the problem is to find the intimate-core group of $Q$.


\begin{example}
In Fig.~\ref{fig.motivate}(a), $G$ is a weighted graph with 12 nodes and 20 edges. Each edge has a positive weight. Given two query nodes $Q = \{ v_{8}, v_{10} \}$ and $k = 3$, the answer of intimate-core group for $Q$ is the subgraph shown in Fig.~\ref{fig.motivate}(b). This is a connected 3-core, and  also containing two query nodes $\{ v_{8}, v_{10} \}$. Moreover, it has the minimum group weight among all connected 3-core subgraphs containing $Q$. 
\end{example}

\subsection{Existing Intimate-Core Group Search Algorithms}
The problem of intimate-core group search has been studied in the literature~\cite{zheng2017querying}. Two heuristic algorithms, namely, \icgs and \icgm, are proposed to deal with this problem in an online manner. No optimal algorithms have been proposed yet because this problem has been proven to be \NPhard~\cite{zheng2017querying}. The NP-hardness is shown by reducing the NP-complete clique decision problem to the intimate-core group search problem. 

Existing solutions \icgs and \icgm both first identify a maximal connected $k$-core as a candidate, and then remove the node with the largest weight of its incident edges at each iteration~\cite{zheng2017querying}. The difference between \icgs and \icgm lies on the node removal. \icgs removes one node at each iteration, while \icgm removes a batch of nodes at each iteration. Although \icgm can significantly reduce the total number of removal iterations required by \icgs, it still takes a large number of iterations for large networks. The reason is that the initial candidate subgraph connecting all query nodes is the maximal connected $k$-core, which may be too large to shrink. This, however, is not always necessary. In particular, if there exists a small connected $k$-core surrounding query nodes, then a few numbers of iterations may be enough token for finding answers. This paper proposes a local exploration algorithm to find a smaller candidate subgraph. On the other hand, both \icgs and \icgm apply the core decomposition to identify the $k$-core from scratch, which is also costly expensive. To improve efficiency, we propose to construct an index offline and retrieve $k$-core for queries online.


\section{Index-Based Local Exploration Algorithms} \label{sec.algo}

In this section, we first introduce a useful core index and the index construction algorithm. Then, we present the index-based intimate-core group search algorithms using local exploration. 

\subsection{K-Core Index}

\begin{algorithm}[t]
\small
\caption{Core Index Construction} \label{algo.index}
\begin{flushleft}
\textbf{Input:} A weighted graph $G=(V, E, w)$\\
\textbf{Output:} Coreness $\delta(v)$ for each $v \in V_{G}$\\
\end{flushleft}
\
\begin{algorithmic}[1]

\STATE	Sort all nodes in $G$ in ascending order of their degree;

\STATE	\textbf{while} $G \neq \emptyset$ 

\STATE	\hspace{0.3cm} Let $d$ be the minimum degree in $G$;

\STATE	\hspace{0.3cm} \textbf{while} there exists $deg_{G}(v) \leq d$ 

\STATE	\hspace{0.3cm} \hspace{0.3cm} $\delta(v) \leftarrow d$;

\STATE	\hspace{0.3cm} \hspace{0.3cm} Remove $v$ and its incident edges from $G$;

\STATE	\hspace{0.3cm} \hspace{0.3cm} Re-order the remaining nodes in $G$ in ascending order of their degree;

\STATE  Store $\delta(v)$ in index for each $v \in V_{G}$;

\end{algorithmic}
\end{algorithm}

We start with a useful definition of coreness as follows.

\begin{definition}
[Coreness] The coreness of a node $v\in V$, denoted by $\delta(v)$, is the largest number $k$ such that there exists a connected $k$-core containing $v$.
\end{definition}

Obviously, for a node $q$ with the coreness $\delta(q)= l$, there exists a connected $k$-core containing $q$ where $1\leq k\leq l$; meanwhile, there is no connected $k$-core containing $q$ where $k >l$. The $k$-core index keeps the coreness of all nodes in $G$. 


\stitle{K-core index construction.} We apply the existing core decomposition~\cite{batagelj2003m} on graph $G$ to construct the $k$-core index. The algorithm is outlined in Algorithm~\ref{algo.index}. The core decomposition is to compute the coreness of each node in graph $G$. Note that for the self-completeness of our techniques and reproducibility, the detailed algorithm of core decomposition is also presented (lines 1-7).
First, the algorithm sort all nodes in $G$ based on their degree in ascending order. Second, it finds the minimum degree in $G$ as $d$. Based on the definition of $k$-core, it next computes the coreness of nodes with $deg_{G}(v)=d$ as $d$ and removing these nodes and their incident edges from $G$. With the deletion of these nodes, the degree of neighbors of these nodes will decrease. For those nodes which have a new degree at most $d$, they will not be in (d+1)-core while they will get $\delta(v)=d$. It continues the removal of nodes until there is no node has $deg_{G}(v) \leq d$. Then, the algorithm back to line 2 and starts a new iteration to compute the coreness of remaining nodes.
Finally, it stores the coreness of each vertex $v$ in $G$ as the $k$-core index.

\subsection{Solution Overview}

\begin{algorithm}[t]
\small
\caption{\leks Framework} \label{algo.fm}
\begin{flushleft}
\textbf{Input:} $G=(V, E, w)$, an integer $k$, a set of query vertices $Q$\\
\textbf{Output:} Intimate-core group $H$\\
\end{flushleft}
\
\begin{algorithmic}[1]


\STATE	Find a tree $T_Q$ for query nodes $Q$ using Algorithm~\ref{algo.tree} or Algorithm~\ref{algo.path};

\STATE	Expand the tree $T_Q$ to a candidate graph $G_{Q}$ in Algorithm~\ref{algo.expand};

\STATE	Apply \icgm~\cite{zheng2017querying} on graph $G_{Q}$;

\STATE	Return a refined intimate-core group as answers;  


\end{algorithmic}
\end{algorithm}

\begin{figure}[!h]
     \includegraphics[width=\textwidth]{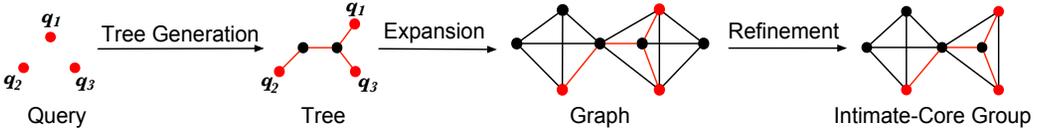}
     \caption{\leks framework for intimate-core group search}
     \label{fig.fm}
\end{figure}

At a high level, our algorithm of \underline{l}ocal \underline{e}xploration based on \underline{k}-core index for intimate-core group \underline{s}earch (\leks) consists of three phases:

\begin{enumerate} 

\item \emph{Tree Generation Phase}: This phase invokes the shortest path algorithm to find the distance between any pair of nodes, and then constructs a small-weighted tree by connecting all query nodes.

\item \emph{Expansion Phase}: This phase expands a tree into a graph. It applies the idea of local exploration to add nodes and edges. Finally, it obtains a connected $k$-core containing all query nodes.   

\item \emph{Intimate-Core Refinement Phase}: This phase removes nodes with large weights, and maintains the candidate answer as a connected $k$-core. This refinement process stops until an intimate-core group is obtained.  

\end{enumerate}

Fig.~\ref{fig.fm} shows the whole  framework of our index-based local exploration algorithm. Note that we compute the $k$-core index offline and apply the above solution of online query processing for intimate-core group search.  In addition, we consider $|Q|\geq 2$ for tree generation phase, and skip this phase if $|Q|=1$. Algorithm~\ref{algo.fm} also depicts our algorithmic framework of \leks. 





\subsection{Tree Generation}


\begin{algorithm}[t]
\small
\caption{Tree Construction} \label{algo.tree}
\begin{flushleft}
\textbf{Input:} $G=(V, E, w)$, an integer $k$, a set of query vertices $Q$, the $k$-core index\\
\textbf{Output:} Tree $T_Q$\\
\end{flushleft}
\
\begin{algorithmic}[1]



\STATE  Identify the maximal connected $k$-core of $C_k$ containing query nodes $Q$;

\STATE  Let $G_{pw}$ be an empty graph;

\STATE	\textbf{for} $q_{1}, q_{2} \in Q$ 

\STATE	\hspace{0.3cm} \textbf{if} there is no path between $q_{1}$ and $q_{2}$ in $C_{k}$ \textbf{then}

\STATE	\hspace{0.3cm} \hspace{0.3cm} \textbf{return} $\emptyset$;

\STATE	\hspace{0.3cm} \textbf{else}

\STATE	\hspace{0.3cm} \hspace{0.3cm} Compute the shortest path between $q_1$ and $q_2$  in $C_{k}$;

\STATE	\hspace{0.3cm} \hspace{0.3cm} Add the $\spath_{C_k}(q_1, q_2)$ between $q_1$ and $q_2$ into $G_{pw}$;


\STATE	Tree: $T_Q \leftarrow \emptyset$; 

\STATE	Priority queue: $L \leftarrow \emptyset$;

\STATE	\textbf{for} each node $v$ in $G_{pw}$

\STATE	\hspace{0.3cm} $\dist(v)\leftarrow \infty$;

\STATE	$Q \leftarrow Q -\{q_0\}$; \dist$(q_{0})\leftarrow 0$; $L.push(q_{0}, \dist(q_{0}))$;

\STATE	\textbf{while} $Q \neq \emptyset$ \textbf{do}


\STATE	\hspace{0.3cm} Extract a node $v$ and its edges with the smallest $\dist(v)$ from $L$;



\STATE	\hspace{0.3cm} Insert node $v$ and its edges into $T_Q$;

\STATE	\hspace{0.3cm} \textbf{if} $v \in Q$  \textbf{then}

\STATE	\hspace{0.3cm} \hspace{0.3cm} $Q \leftarrow Q -\{v\}$;

\STATE	\hspace{0.3cm} \textbf{for} $u \in N_{G_{pw}}(v)$ \textbf{do}

\STATE	\hspace{0.3cm} \hspace{0.3cm} \textbf{if} $\dist(u) > w(u, v)$ \textbf{then}

\STATE	\hspace{0.3cm} \hspace{0.3cm} \hspace{0.3cm} $\dist(u) \leftarrow w(u, v)$;  

\STATE	\hspace{0.3cm} \hspace{0.3cm} \hspace{0.3cm} Update  $(u, \dist(u))$ in $L$;

\STATE    \textbf{return} $T_Q$;



\end{algorithmic}
\end{algorithm}

In this section, we present the phase of tree generation. Due to the large-scale size of $k$-core in practice, we propose local exploration methods to identify small-scale substructures as candidates from the $k$-core. The approaches produce a tree structure with small weights to connect all query nodes. We develop two algorithms, respectively based on the minimum spanning tree (\MST) and minimum weighted path (\MWP). 



\stitle{Tree construction.} The tree construction has three major steps. Specifically, the algorithm firstly generates all-pairs shortest paths for query nodes $Q$ in the $k$-core $C_k$ (lines 1-7).  Given a path between nodes $u$ and $v$, the path weight  is the total weight of all edges along this path between $u$ and $v$.  It uses \spath$_{\mathcal{C}_k}(u,v)$ to represent the shortest path between nodes $u$ and $v$ in the $k$-core $C_k$. For any pair of query nodes $q_i$, $q_j \in Q$, our algorithm invokes the well-known Dijkstra's algorithm~\cite{cormen2009introduction} to find the shortest path \spath$_{C_k}(q_i,q_j)$ in the $k$-core $C_k$.

Second, the algorithm constructs a weighted graph $G_{pw}$ for connecting all query nodes (lines 3-8). Based on the obtained all-pairs shortest paths, it collects and merges all these paths together to construct a weighted graph $G_{pw}$ correspondingly. 

Third, the algorithm generates a small spanning tree for $Q$ in the weighted graph $G_{pw}$ (lines 9-22), since not all edges are needed to keep the query nodes connected in $G_{pw}$. This step finds a compact spanning tree to connect all query nodes $Q$, which removes no useful edges to reduce weights. Specifically, the algorithm starts from one of the query nodes and does expand based on Prim's minimum spanning tree algorithm~\cite{cormen2009introduction}. The algorithm stops when all query nodes are connected into a component in $G_{pw}$. Against the maximal connected $k$-core, our compact spanning tree has three significant features: (1) Query-centric. The tree involves all query nodes of $Q$. (2) Compactly connected. The tree is a connected and compact structure; (3) Small-weighted. The generation of minimum spanning tree ensures a small weight of the discovered tree.

\begin{figure}[ht]
	 \begin{tabular} {@{\hspace{0.12\textwidth}}c@{\hspace{0.12\textwidth}} c@{\hspace{0.1\textwidth}}}
     \centering
     	\includegraphics[width=0.25\textwidth]{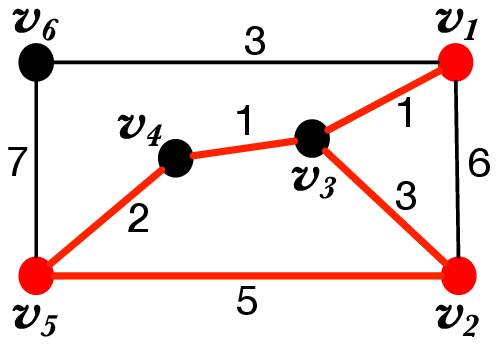} &
     	\includegraphics[width=0.25\textwidth]{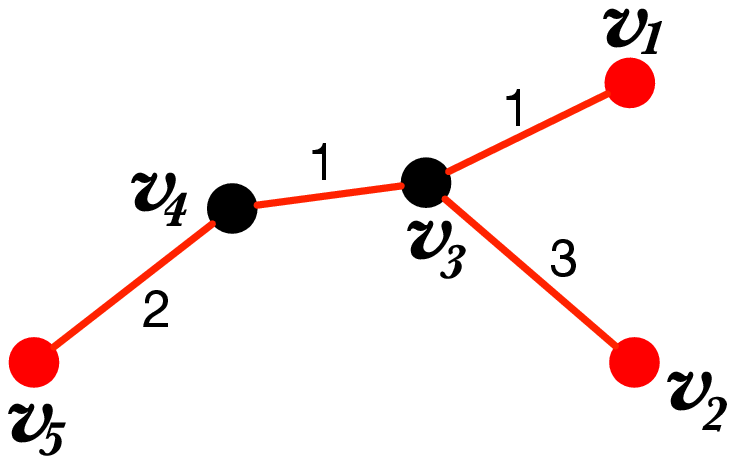} \\
     	\newline
     	(a) Find all pairs of shortest path & (b) Spanning tree\\
     \end{tabular}
     \caption{Tree construction for query nodes $v_{1}$, $v_{2}$, $v_{5}$.}\label{fig.tree}
     
\end{figure}

\begin{example}
	Fig.~\ref{fig.tree}(a) shows a weighted graph $G$ with 6 nodes and 8 edges with weights. Assume that $k=2$, the whole graph is 2-core as $C_2$. A set of query nodes $Q = \{ v_{1}, v_{2}, v_{5}\}$ are colored in red in Fig.~\ref{fig.tree}(a). We first find the shortest path between every pair of query nodes in $Q$. All edges along with these shortest path are colored in red in Fig.~\ref{fig.tree}(a). For example, the shortest path between $v_1$ and $v_2$ is  $\spath_{C_2}(v_{1}, v_{2}) = \{(v_{1}, v_{3}), (v_3, v_2)\}$. Similarly, $\spath_{C_2}$$(v_{1}, v_{5}) $ $= \{(v_1, v_3), (v_3, v_4), (v_4, v_5)\}$, $\spath_{C_2}$ $(v_{2}, v_{5}) = \{(v_{2}, v_{5} )\}$. All three paths are merged to construct a weighted graph $G_{pw}$ in red in Fig.~\ref{fig.tree}(a). A spanning tree of $T_{Q}$ is shown in Fig.~\ref{algo.tree}(b), which connects all query nodes $\{v_{1}, v_{2}, v_{5}\}$ with a small weight of 7.
\end{example}

\begin{algorithm}[t]
\small
\caption{Path-based Construction} \label{algo.path}
\begin{flushleft}
\textbf{Input:} $G=(V, E, w)$, an integer $k$, a set of query vertices $Q$, the $k$-core index\\
\textbf{Output:} Tree $T_{Q}$\\
\end{flushleft}
\
\begin{algorithmic}[1]

\STATE    Identify the maximal connected $k$-core of $C_k$ containing query nodes $Q$;

\STATE	Let $q_{0}$ be the first query node of $Q$;

\STATE	$Q \leftarrow Q - \{q_{0}\}$;

\STATE	\textbf{while} $Q \neq \emptyset$ \textbf{do}

\STATE	\hspace{0.3cm} \textbf{if} there is no path between $q$ and $q_{0}$ in $C_{k}$ \textbf{then}

\STATE	\hspace{0.3cm} \hspace{0.3cm} \textbf{return} $\emptyset$;

\STATE	\hspace{0.3cm} \textbf{else}

\STATE	\hspace{0.3cm} \hspace{0.3cm} Compute the shortest path between $q$ and $q_0$  in $C_{k}$;

\STATE	\hspace{0.3cm} \hspace{0.3cm} Add the $\spath_{C_k}(q, q_0)$ between $q$ and $q_0$ into $T_{Q}$;

\STATE	\hspace{0.3cm} \hspace{0.3cm} $q_0 \leftarrow q$, $Q \leftarrow Q - \{q_0\}$;

\STATE  \textbf{return} $T_{Q}$;

\end{algorithmic}
\end{algorithm}

\stitle{Path-based construction.} Algorithm~\ref{algo.tree} may take expensive computation for finding the shortest path between every pair of nodes that are far away from each other. To improve efficiency, we develop a path-based approach to connect all query nodes directly. The path-based construction is outlined in Algorithm~\ref{algo.path}. The algorithm starts from one query node $q_0$, and searches the shortest path to the nearest query node in $Q$(lines 2-8). After that, it collects and merges the weighted path $\spath_{C_k}(q, q_0)$ into $T_{Q}$ to construct the tree(line 9). Recursively, it starts from the new query node $q$ as $q_0$ to find the next nearest query node $q$, until all query nodes in $Q$ are found in such a way(line 10). The algorithm returns the tree connecting all query nodes. 

\begin{example} We apply Algorithm~\ref{algo.path} on graph $G$ in Fig.~\ref{fig.tree}(a) with query $Q = \{ v_{1}, v_{2}, v_{5}\}$ and $k=2$. We start the shortest path search from $v_{1}$. The nearest query node to $v_1$ is $v_5$, we can find the shortest path $\spath_{C_2}(v_{1}, v_{5})$ $ = \{ (v_1, v_3), (v_3, v_4), (v_4, v_5) \}$. Next, we start from $v_{5}$ and find the shortest path $\spath_{C_2}$ $(v_{5}, v_{2}) $ $= \{(v_{5}, v_{2}) \}$. Finally, we merge the two paths $\spath_{C_2}(v_{1}, v_{5})$ and $\spath_{C_2}(v_{5}, v_{2})$ to construct the tree $T_{Q}$.
\end{example}

\stitle{Complexity analysis.} We analyze the complexity of Algorithm~\ref{algo.tree} and Algorithm~\ref{algo.path}. Assume that the $k$-core $C_k$  has $n_k$ nodes and $m_k$ edges where $n_k\leq n$ and $m_k\leq m$. 

For Algorithm~\ref{algo.tree}, an intuitive implementation of all-pairs-shortest-paths needs to compute the shortest path for every pair nodes in $Q$, which takes $O(|Q|^2m_k\log n_k)$ time. However, a fast implementation of single-source-shortest-path algorithm  can compute the shortest path from one query node $q\in Q$ to all other nodes in $Q$, which takes $O(m_k\log n_k)$ time. Overall, the computation of all-pairs-shortest-paths can be done in $O(|Q|m_k\log n_k)$ time. In addition, the weighted graph $G_{pw}$ is a subgraph of $C_k$, thus the size of $G_{pw}$ is $O(n_k+m_k) \subseteq O(m_k)$. Identifying the spanning tree of $G_{pw}$ takes $O(m_k\log n_k)$ time. Overall, Algorithm~\ref{algo.tree} takes $O(|Q|m_k\log n_k)$ time and $O(m_k)$ space. 

For Algorithm~\ref{algo.path}, it applies $|Q|$ times of single-source-shortest-path to identify the nearest query node. Thus, Algorithm~\ref{algo.path} also takes $O(|Q|m_k\log n_k)$ time and $O(m_k)$ space. In practice, Algorithm~\ref{algo.path} runs faster than Algorithm~\ref{algo.tree} on large real-world graphs, which avoids the weighted tree construction and all-pairs-shortest-paths detection.



\subsection{Tree-to-Graph Expansion}

\begin{algorithm} [t]
\small
\caption{Tree-to-Graph Expansion} \label{algo.expand}
\begin{flushleft}
\textbf{Input:} $G=(V, E, w)$, a set of query vertices $Q$, $k$-core index, $T_Q$\\
\textbf{Output:} Candidate subgraph $G_{Q}$\\
\end{flushleft}
\
\begin{algorithmic}[1]

\STATE  Identify the maximal connected $k$-core of $C_k$ containing query nodes $Q$;

\STATE	$L_{0} \leftarrow \{ v | v \in V_{T_Q} \}$; $L' \leftarrow L_{0}$ ;

\STATE  $i \leftarrow 0$; $G_{Q} \leftarrow \emptyset$;

\STATE	\textbf{while} $G_{Q} = \emptyset$ \textbf{do}

\STATE	\hspace{0.3cm} \textbf{for} each $v \in L_{i}$ \textbf{do}

\STATE	\hspace{0.3cm} \hspace{0.3cm} \textbf{for} each $u \in N_{C_{k}}(v)$ and $u \notin L' \cup L_{i+1}$ \textbf{do}

\STATE	\hspace{0.3cm} \hspace{0.3cm} \hspace{0.3cm} $L_{i+1} \leftarrow L_{i+1} \cup \{u\}$;

\STATE	\hspace{0.3cm} $L' \leftarrow L' \cup L_{i+1}$; $i \leftarrow i+1$;

\STATE	\hspace{0.3cm} 
Let $G_L$ be the induced subgraph of $G$ by the node set $L'$; 

\STATE	\hspace{0.3cm} Generate a connected $k$-core of $G_{L}$ containing query nodes $Q$ as $G_{Q}$;

\STATE  \textbf{return} $G_{Q}$;

\end{algorithmic}
\end{algorithm}

In this section, we introduce the phase of tree-to-graph expansion. This method expands the obtained tree from Algorithm~\ref{algo.tree} or Algorithm~\ref{algo.path} into a connected $k$-core candidate subgraph $G_{Q}$. It consists of two main steps. First, it adds nodes/edges to expand the tree into a graph layer by layer. Then, it prunes disqualified nodes/edges to maintain the remaining graph as a connected $k$-core. The whole procedure is shown in Algorithm~\ref{algo.expand}. 

Algorithm~\ref{algo.expand} first gets all nodes in $T_{Q}$ and puts them into $L_0$ (line 2). Let $L_{i}$ be the vertex set at the $i$-th depth of expansion tree, and $L_{0}$ be the initial set of vertices. It uses $L'$ to represent the set of candidate vertices, which is the union of all $L_{i}$ set. The iterative procedure can be divided into three steps (lines 4-10). First, for each vertex $v$ in $L_{i}$, it adds their neighbors into $L_{i+1}$ (lines 5-7). Next, it collects and merges $\{ L_0, ..., L_{i+1} \}$ into $L'$ and constructs a candidate graph $G_L$  as the induced subgraph of $G$ by the node set $L'$ (lines 8-9). 
Finally, we apply the core decomposition algorithm on $G_L$ to find the connected $k$-core subgraph containing all query nodes, denoted as $G_Q$. If there exists no such $G_Q$, Algorithm~\ref{algo.expand} explores the $(i+1)$-th depth of expansion tree and repeats the above procedure (lines 4-10). In the worst case, $G_Q$ is exactly the maximum connected $k$-core subgraph containing $Q$. However, $G_Q$ in practice is always much smaller than it. The time complexity for expansion is $O(\sum_{i=0}^{l_{max}} \sum_{v\in V(G_i)} \deg(v) )$, where $l_{max}$ is the iteration number of expansion in Algorithm~\ref{algo.expand}.   

\begin{figure}[ht] 
     \begin{tabular} {@{\hspace{0.1\textwidth}}c @{\hspace{0.1\textwidth}} c@{\hspace{0.1\textwidth}}}
     \centering
     	\hfil\includegraphics[width=0.32\textwidth]{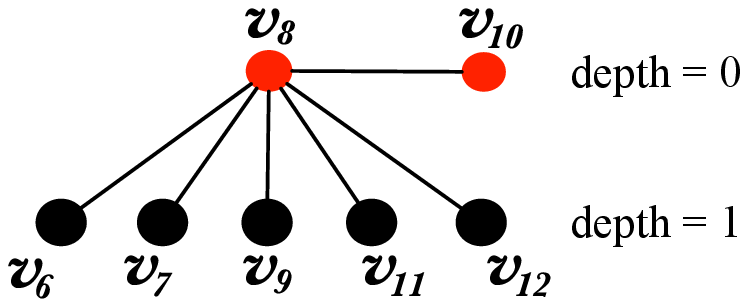} &
     	\hfil\includegraphics[width=0.32\textwidth]{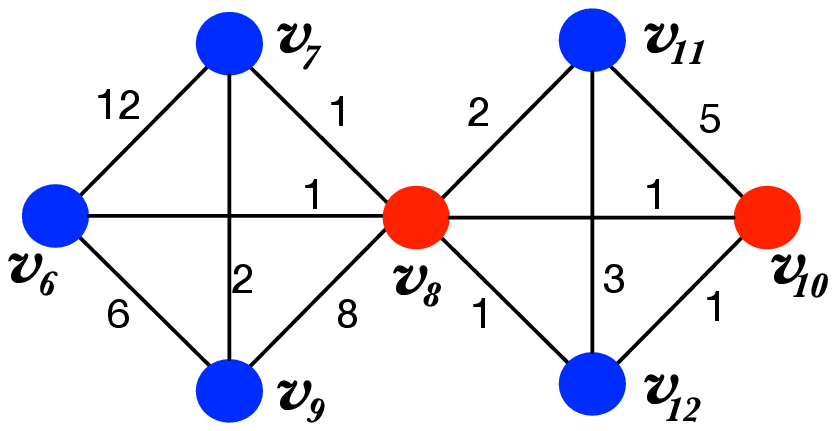} \\
     	\newline
     	(a) Expansion & (b) Candidate subgraph construction\\
     \end{tabular}
     \caption{Tree-to-graph expansion}\label{fig.expand}
\end{figure}

\begin{example} Fig.~\ref{fig.motivate}(a) shows a weighted graph $G$ with query $Q = \{ v_{8}, v_{10} \}$ and $k=3$. We first identify the maximal connected 3-core  containing query nodes $Q$. Since there is only 2 query nodes, the spanning tree is same as the shortest path between them, such that $T_{Q}=\spath_{C_3} (v_{8}, v_{10})$. Next, we initialize $L_0$ as $L_0 = \{v_{8}, v_{10}\}$ and expand nodes in $L_0$ to their neighbors. The expansion procedure is shown in Fig.~\ref{fig.expand}(a). We put all nodes in Fig.~\ref{fig.expand}(a) into $L'$ and construct a candidate subgraph $G_L$ shown in Fig.~\ref{fig.expand}(b). Since $G_L$ is a 3-core connected subgraph containing query nodes, the expansion graph $G_Q$ is $G_L$ itself.
\end{example}

\subsection{Intimate-Core Refinement}
This phase refines the candidate connected $k$-core into an answer of the intimate-core group. We apply the existing approach \icgm~\cite{zheng2017querying} by removing nodes to shrink the candidate graph obtained from Algorithm~\ref{algo.expand}.  
This step takes $O(m' \log_{\varepsilon}n')$ time, where $\varepsilon > 0$ is a parameter of shrinking graph~\cite{zheng2017querying}. To avoid query nodes deleted by the removal processes of \icgm, we develop a mechanism to protect important query nodes. 

\stitle{Protection mechanism for query nodes.} As pointed by~\cite{zhang2017finding,bhawalkar2015preventing,zhang2017olak}, the $k$-core structure may collapse when critical nodes are removed. Thus, we precompute such critical nodes for query nodes in $k$-core and ensure that they are not deleted in any situations. We use an example to illustrate our ideas. For a query node $q$ with an exact degree of $k$, it means that if any neighbor is deleted, there exists no feasible $k$-core containing $q$ any more. Thus, $q$ and all $q$'s neighbors are needed to protect. For example, in Fig.~\ref{fig.expand}(b), assume that $k=3$, there exists $deg_{G}(v_{10})=k$. The removal of each node in $N_{G}(v_{10})$ will cause core decomposition and the deletion of $v_{10}$. 
This protection mechanism for query nodes can also be used for $k$-core maintenance in the phrase of tree-to-graph expansion.



\section{Experiments} \label{sec.exp}

\begin{figure} [t]
     \begin{tabular} {@{\hspace{0.02\textwidth}}c@{\hspace{0.02\textwidth}} c@{\hspace{0.02\textwidth}} c}
     \centering
          \hfil\includegraphics[width=0.32\textwidth]{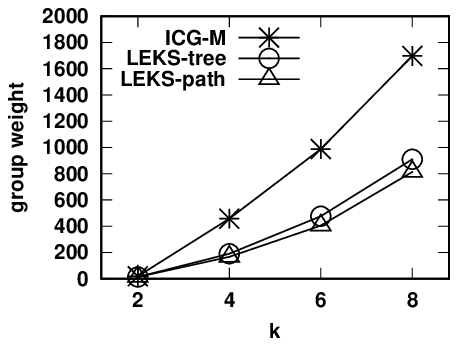} &
          \hfil\includegraphics[width=0.32\textwidth]{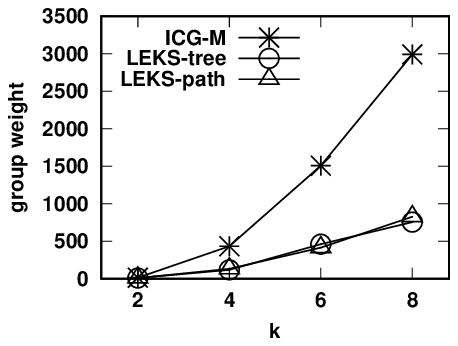} &
          \hfil\includegraphics[width=0.32\textwidth]{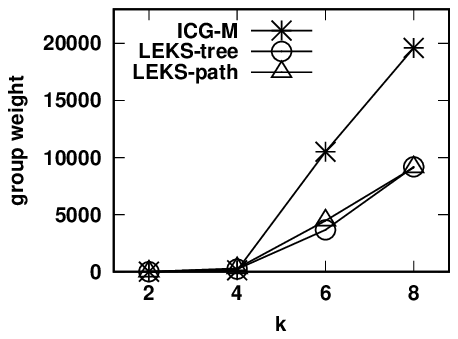} \\
          \newline
          (a) wiki-vote & (b) Flickr & (c) DBLP\\
     \end{tabular}
     \caption{Effectiveness evaluation by varying k} \label{fig.quality_k}
\end{figure}

\begin{figure}  [t]
     \begin{tabular} {@{\hspace{0.02\textwidth}}c@{\hspace{0.02\textwidth}} c@{\hspace{0.02\textwidth}} c}
     \centering
          \hfil\includegraphics[width=0.32\textwidth]{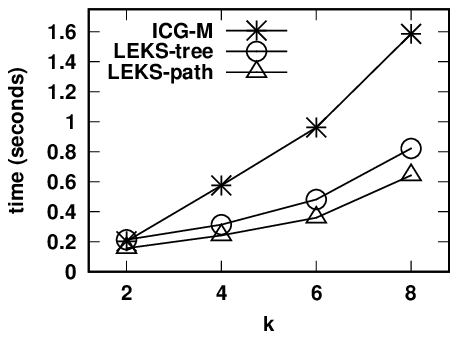} &
          \hfil\includegraphics[width=0.32\textwidth]{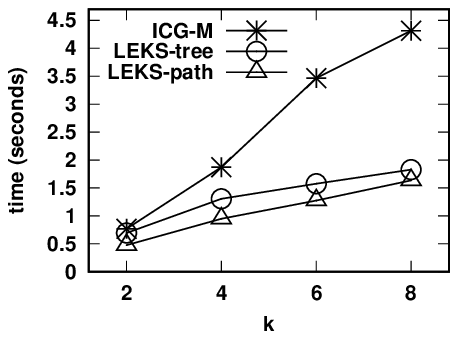} &
          \hfil\includegraphics[width=0.32\textwidth]{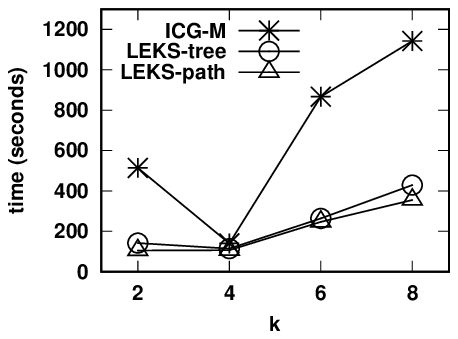}\\
          \newline
          (a) wiki-vote & (b) Flickr & (c) DBLP\\
     \end{tabular}
     \caption{Efficiency evaluation by varying k}\label{fig.time_k}
\end{figure}

In this section, we experimentally evaluate the performance of our proposed algorithms. All algorithms are implemented in Java and performed on a Linux server with Xeon E5-2630 (2.2 GHz) and 256 GB RAM.

\stitle{Datasets.} We use three real-world datasets in experiments. All datasets are publicly available from~\cite{huang2016truss}. The edge weight represents the existence probability of an edge. A smaller weight indicates a higher possibility of the edge to existing. The statistics of datasets are shown in Table~\ref{tab.dataset}. The maximum coreness $\delta_{max} =\max_{v\in V} \delta(v)$.

\begin{table}[h] 
\centering
\caption{Network statistics}\label{tab.dataset}
\begin{tabular}{|l|l|l|l|}
\hline
Datasets & $|V|$ & $|E|$ & $\delta_{max}$ \\
\hline
wiki-vote & 7,115 & 103,689 & 56 \\
Flickr & 24,125 & 300,836 & 225 \\
DBLP & 684,911 & 2,284,991 & 114 \\
\hline
\end{tabular}
\end{table}

\stitle{Algorithms.} We compare 3 algorithms as follows. 

\squishlisttight
\item \icgm: is the state-of-the-art approach for finding intimate-core group using  bulk deletion~\cite{zheng2017querying}.
\item \ltree: is our index-based search framework in Algorithm~\ref{algo.fm} using Algorithm~\ref{algo.tree} for tree generation. 
\item \lpath: is our index-based search framework in Algorithm~\ref{algo.fm} using Algorithm~\ref{algo.path} for tree generation. 
\end{list} 


We evaluate all algorithms by comparing the running time and the intimate-core group weight. The less running time costs, the more efficient the algorithm is. Smaller the group weight of the answer, better effectiveness is. 

\stitle{Queries and parameters.} We evaluate all competitive approaches by varying parameters $k$ and $|Q|$. The range of $k$ is \{2, 4, 6, 8\}. The number of query nodes $|Q|$ falls in \{1, 2, 3, 4, 5, 6, 7\}. We randomly generate 100 sets of queries by different $k$ and $|Q|$. 


\stitle{Exp-1: Varying $k$.} Fig.~\ref{fig.quality_k} shows the group weight of three algorithms by varying parameter $k$ on all datasets. The results show that our local search methods \ltree and \lpath can find intimate groups with lower group weights than \icgm, for different $k$. The performance of \ltree and \lpath are similar. 
Fig.~\ref{fig.time_k} shows that \lpath performs the best for most cases, and runs significantly faster than \icgm. Interestingly, \icgm can find answers quickly for $k=4$, which achieves similar performance with \leks methods.

\begin{figure} [ht]
     \begin{tabular} {@{\hspace{0.02\textwidth}}c@{\hspace{0.02\textwidth}} c@{\hspace{0.02\textwidth}} c}
     \centering
          \hfil\includegraphics[width=0.32\textwidth]{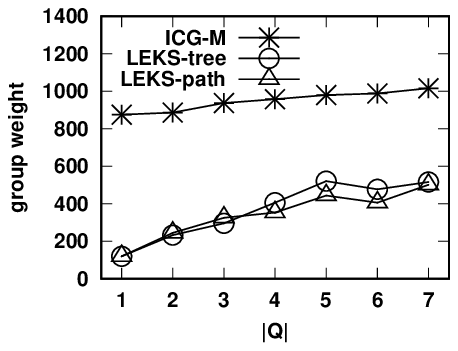} &
          \hfil\includegraphics[width=0.32\textwidth]{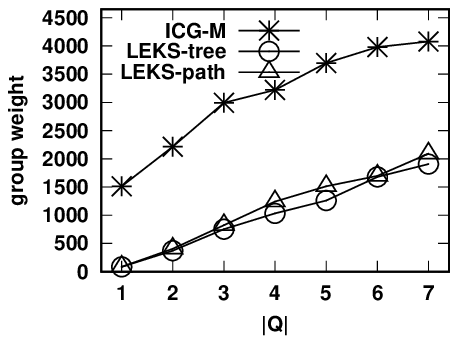} &
          \hfil\includegraphics[width=0.32\textwidth]{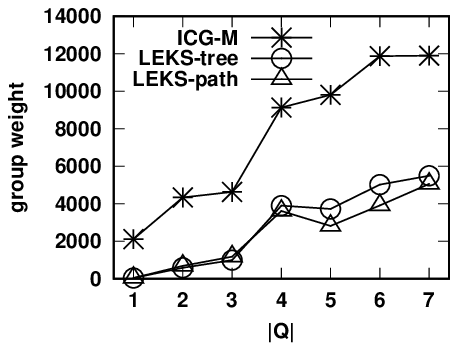}\\
          \newline
          (a) wiki-vote & (b) Flickr & (c) DBLP\\
     \end{tabular}
     \caption{Effectiveness evaluation by varying $|Q|$}\label{fig.quality_q}
\end{figure}

\begin{figure} [ht]
     \begin{tabular} {@{\hspace{0.02\textwidth}}c@{\hspace{0.02\textwidth}} c@{\hspace{0.02\textwidth}} c}
     \centering
          \hfil\includegraphics[width=0.32\textwidth]{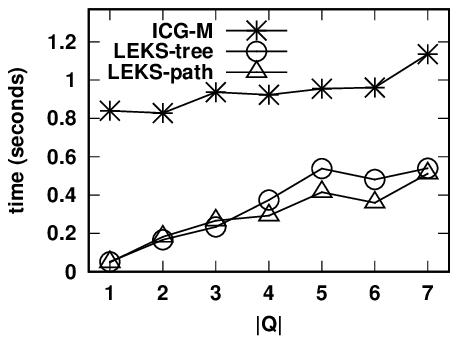} &
          \hfil\includegraphics[width=0.32\textwidth]{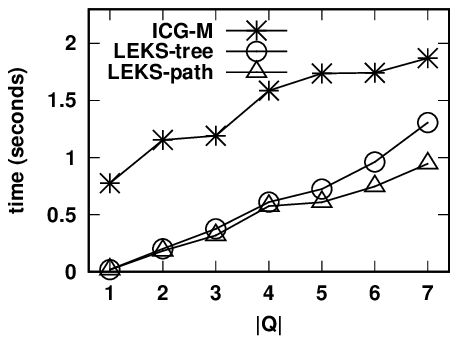} &
          \hfil\includegraphics[width=0.32\textwidth]{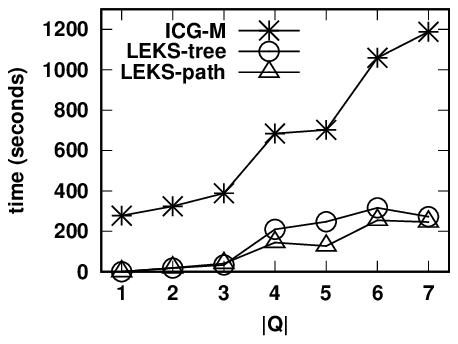}\\
          \newline
          (a) wiki-vote & (b) Flickr & (c) DBLP\\
     \end{tabular}
     \caption{Efficiency evaluation by varying $|Q|$}\label{fig.time_q}
\end{figure}

\stitle{Exp-2: Varying $|Q|$.} Fig.~\ref{fig.quality_q} reports the group weight results of three algorithms for different queries by varying $|Q|$. With the increasing $|Q|$, \ltree and \lpath methods can always find intimate groups with smaller weights than \icgm. \ltree and \lpath perform similarly. Fig.~\ref{fig.time_q} reports the results of running time. It shows that our methods are always faster than \icgm.

\begin{figure}[ht]
     \begin{tabular} {@{\hspace{0.07\textwidth}}c @{\hspace{0.05\textwidth}}c}
     \centering
          \hfil\includegraphics[width=0.38\textwidth]{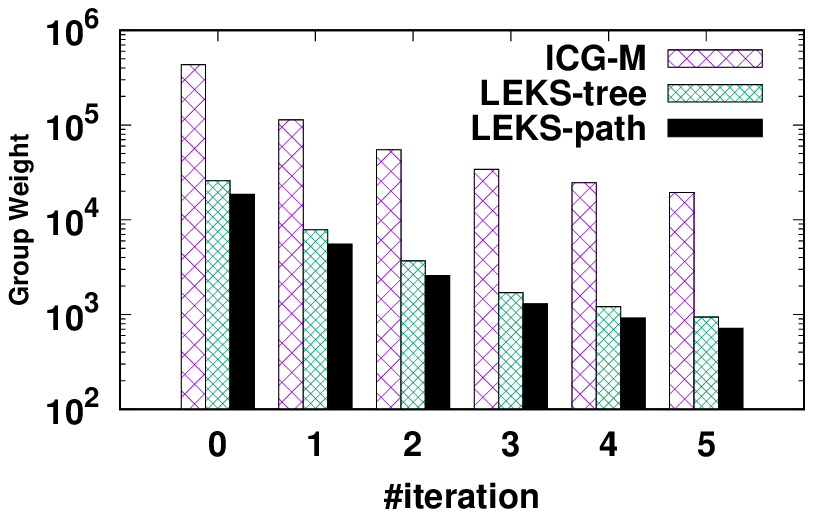} &
          \hfil\includegraphics[width=0.38\textwidth]{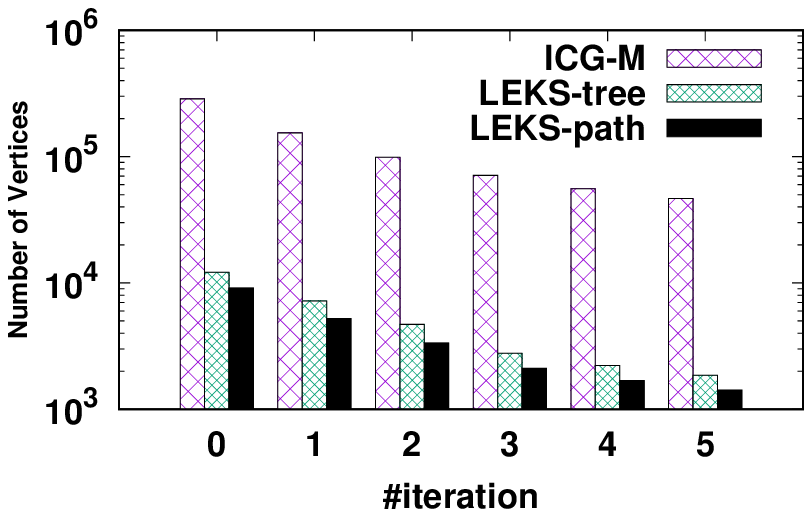}\\
          \newline
          (a) Group weight varied by \#iterations & (b) Number of vertices varied by \#iterations \\
     \end{tabular}
     \caption{The size and weight of intimate-groups varied by \#iterations}\label{fig.iteration}
\end{figure}

\stitle{Exp-3: Quality evaluation of candidate intimate-core groups.} 
This experiment evaluates the subgraphs of candidate intimate-core groups by all methods, in terms of vertex size and group weight. 
\icgm takes the maximal connected $k$-core subgraph containing query nodes as an initial candidate, and iteratively shrinks it. 
\ltree and \lpath both generate an initial candidate subgraph locally expanded from a tree, and then iteratively shrink the candidate by removing nodes. 
We use $k=6$ and $|Q|=5$.  
We report the results of the first 5 removal iterations and the initial candidate at the \#iteration of 0. Fig.~\ref{fig.iteration}(a) shows that the group weight of candidates by our methods is much smaller than \icgm. Fig.~\ref{fig.iteration}(b) reports the vertex size of all candidates at each iteration. The number of vertices in the candidate group by \ltree and \lpath at the \#iteration of 0, is even less than the vertex size of candidate group by \icgm at the \#iteration of 5.

\begin{figure}[ht]
     \begin{tabular} {@{\hspace{0.05\textwidth}}c@{\hspace{0.03\textwidth}} c}
     \centering
          \hfil\includegraphics[width=0.30\textwidth]{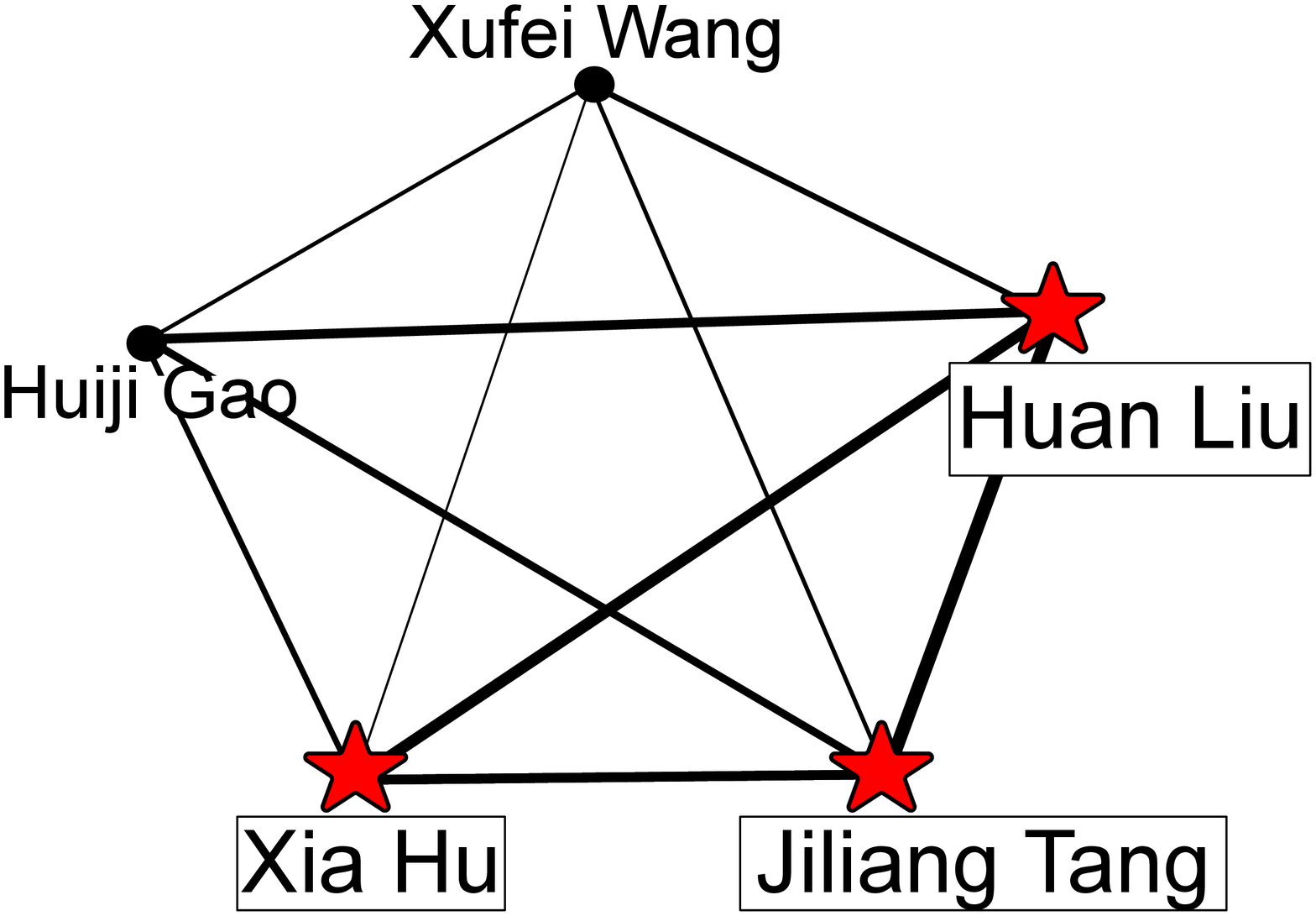} &
          \hfil\includegraphics[width=0.53\textwidth]{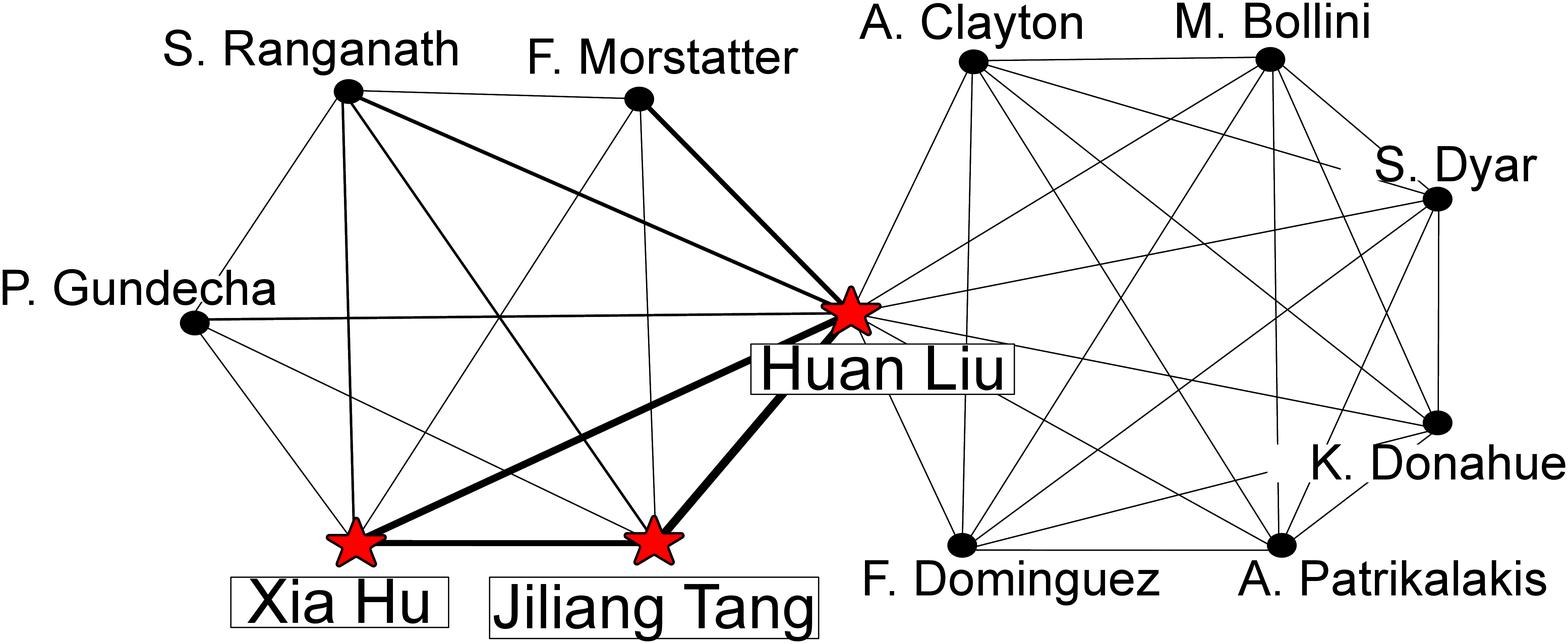}\\
          \newline
          (a)  \lpath & (b)  \icgm \\
     \end{tabular}
     \caption{Case study of intimate-core group search on the DBLP network. Here, query $Q=$\{``Huan Liu", ``Xia Hu", ``Jiliang Tang"\} and $k=4$.}\label{fig.case}
\end{figure}

\stitle{Exp-4: Case study on the DBLP network.} We conduct a case study of intimate-core group search on the collaboration DBLP network~\cite{zheng2017querying}. Each node represents an author, and an edge is added between two authors if they have co-authored papers. The weight of an edge $(u,v)$ is the reciprocal of the number of papers they have co-authored. The smaller weight of $(u,v)$, the closer intimacy between authors $u$ and $v$. We use the query $Q=$\{``Huan Liu", ``Xia Hu", ``Jiliang Tang"\} and $k=4$. We apply \lpath  and \icgm to find 4-core intimate groups for $Q$. The results of  \lpath  and \icgm  are shown in Fig.~\ref{fig.case}(a) and Fig.~\ref{fig.case}(b) respectively. The bolder lines of an edge represent a smaller weight, indicating closer intimate relationships. Our \leks method discovers a compact 4-core with 5 nodes and 10 edges in Fig.~\ref{fig.case}(a), which has the group weight of 1.6, while \icgm finds a subgraph with 12 nodes, which has a larger group weight of 16.7 in Fig.~\ref{fig.case}(b). We can see that nodes on the right side of Fig.~\ref{fig.case}(b) has no co-author connections with two query nodes ``Xia Hu" and ``Jiliang Tang" at all. This case study verifies that our \lpath can successfully find a better intimate-core group than \icgm.

\section{Conclusion} \label{sec.con}
This paper presents a local exploration $k$-core search (\leks) framework for efficient intimate-core group search. \leks generates a spanning tree to connect query nodes in a compact structure, and locally expands it for intimate-core group refinement. Extensive experiments on real datasets show that our approach achieves a higher quality of answers using less running time, in comparison with the existing \icgm method.



\end{document}